\documentclass[10pt, aps, prb, twocolumn, showpacs, superscriptaddress, floatfix]{revtex4-1}

\usepackage{amsmath, amssymb, mathtools, array, color}
\usepackage{hyperref}
\usepackage{graphicx}

\newcommand{ \bra }			[1] { \left<{#1}\right| }
\newcommand{ \ket }			[1] { \left|{#1}\right> }

\newcommand{ \ex }				[1] { \langle{#1}\rangle }

\newcommand{ \ac }				[1] { \left\{{#1}\right\} }

\newcommand{\nn}				{\nonumber}

\newcommand{\at}            			{t_0}
\newcommand{\bt}				{t_{\rm x}}

\def\blfootnote{\xdef\@thefnmark{}\@footnotetext}

\begin{document}


\title{Detecting Majorana Bound States by Nanomechanics}

\author{Stefan Walter}
\affiliation{
Institute for Theoretical Physics and Astrophysics, University of W{\"u}rzburg, 97074 W{\"u}rzburg, Germany
}

\author{Thomas L. Schmidt}
\affiliation{
Department of Physics, Yale University, 217 Prospect Street, New Haven, Connecticut 06520, USA
}

\author{Kjetil B{\o}rkje}
\affiliation{
Department of Physics, Yale University, 217 Prospect Street, New Haven, Connecticut 06520, USA
}

\author{Bj{\"o}rn Trauzettel}
\affiliation{
Institute for Theoretical Physics and Astrophysics, University of W{\"u}rzburg, 97074 W{\"u}rzburg, Germany
}

\date{\today}


\begin{abstract}
We propose a nanomechanical detection scheme for Majorana bound states, which have been predicted to exist at the
edges of a one-dimensional topological superconductor, implemented, for instance, using a semiconducting wire placed
on top of an $s$-wave superconductor. The detector makes use of an oscillating electrode, which can be realized using
a doubly clamped metallic beam, tunnel coupled to one edge of the topological superconductor. We find that a measurement
of the nonlinear differential conductance provides the necessary information to uniquely identify Majorana bound states.
\end{abstract}

\maketitle

\section{Introduction}

Over the past few years, nanomechanics and the field of topological condensed matter systems have been pushing
limits in their respective area of research. Nanomechanical systems have proven to be exceptional measurement devices
for, e.g., mass, force, and position \cite{LaHaye:2004p88,Lassagne:2009p77,Mamin:2001p18,Steele:2009p184,Yang:2006p1},
as well as a unique platform for studying fundamental questions concerning the quantum nature of macroscopic objects,
theoretically as well as experimentally \cite{Marshall:2003p130401,Schmidt:2010p52,Oconnell:2010p97,Teufel:2011p359}.

Majorana fermions are among the most intriguing features of topological states of matter.
They are their own anti-particles, i.e., $\gamma = \gamma^{\dag}$, and satisfy fermionic anticommutation relations
$\ac{\gamma_{i},\gamma_{j}} = 2 \, \delta_{i j}$. A Majorana fermion has half the degrees of freedom of a Dirac fermion. This can
be seen, for instance, by expressing two Majorana fermions $\gamma_{L,R}$ as $\gamma_{R} = c^{\dag}+c$ and
$\gamma_{L} = -i (c^{\dag}-c)$, where $c$ and $c^\dag$ are the annihilation and creation operators, respectively, of a single
Dirac fermion and satisfy $\ac{c^{\dag},c} = 1$. Various proposals have been made about how to generate Majorana states
\cite{Read:2000p256,Kitaev:2001p131,Fu:2008p273,Fu:2009p228,Sau:2010p254,Lutchyn:2010p077001,Oreg:2010p177002,Alicea:2010p125318}.
In view of future applications, these states might be particularly useful in the field of topological quantum computation
\cite{Nayak:2008p51}. Unfortunately, to date no experimental realization of Majorana fermions has been achieved. Proposed detection
schemes are based on tunnel setups \cite{Bolech:2007p237002,Nilsson:2008p120403,Law:2009p237001,Flensberg:2010p180516,Leijnse:2010a1012.4650v1},
interferometer setups \cite{Akhmerov:2009p233,Fu:2009p257,Fu:2010p258,Beri:2011p263,Liu:2011p262} and the Josephson
effect \cite{Fu:2008p273,Fu:2009p228,Lutchyn:2010p077001}. However it is fair to say that true qualitative experimental signatures
of Majorana fermions that persist in realistic systems are rather difficult to predict.

In this paper, we present a novel scheme for detecting Majorana bound states (MBS) at the edges of topological superconductors.
Our proposal involves an oscillating electrode tunnel coupled to a topological superconductor (TS) hosting MBS
(see Fig.~\ref{fig:modelsystem}). We show that the interplay between the two weakly coupled MBS, located at the edges of the TS,
and the oscillating electrode gives rise to unique features in the differential conductance of the setup. We find that in the
presence of the resonator, satellite peaks appear in the differential conductance. We identify the underlying transport
processes giving rise to the rich structure in the differential conductance. Furthermore, we study the dependence of
the differential conductance on the effective temperature of the oscillator and find for an oscillator close to its quantum
mechanical ground state a qualitative transport signature which is due to the interplay between the resonator and
the Majorana bound states.

The paper is organized as follows. We introduce the setup under investigation and the underlying model in
Sec.~\ref{sec:setupmodel}, followed by details on the calculation of the current in the setup in Sec.~\ref{sec:current}. In
Sec.~\ref{sec:results}, we present our results with subsections focusing on a setup with and without the
resonator. Finally, we conclude in Sec.~\ref{sec:conclusion}.

\section{Setup and Model}
\label{sec:setupmodel}

\begin{figure}[t]
	\center{\includegraphics[width=1\columnwidth]{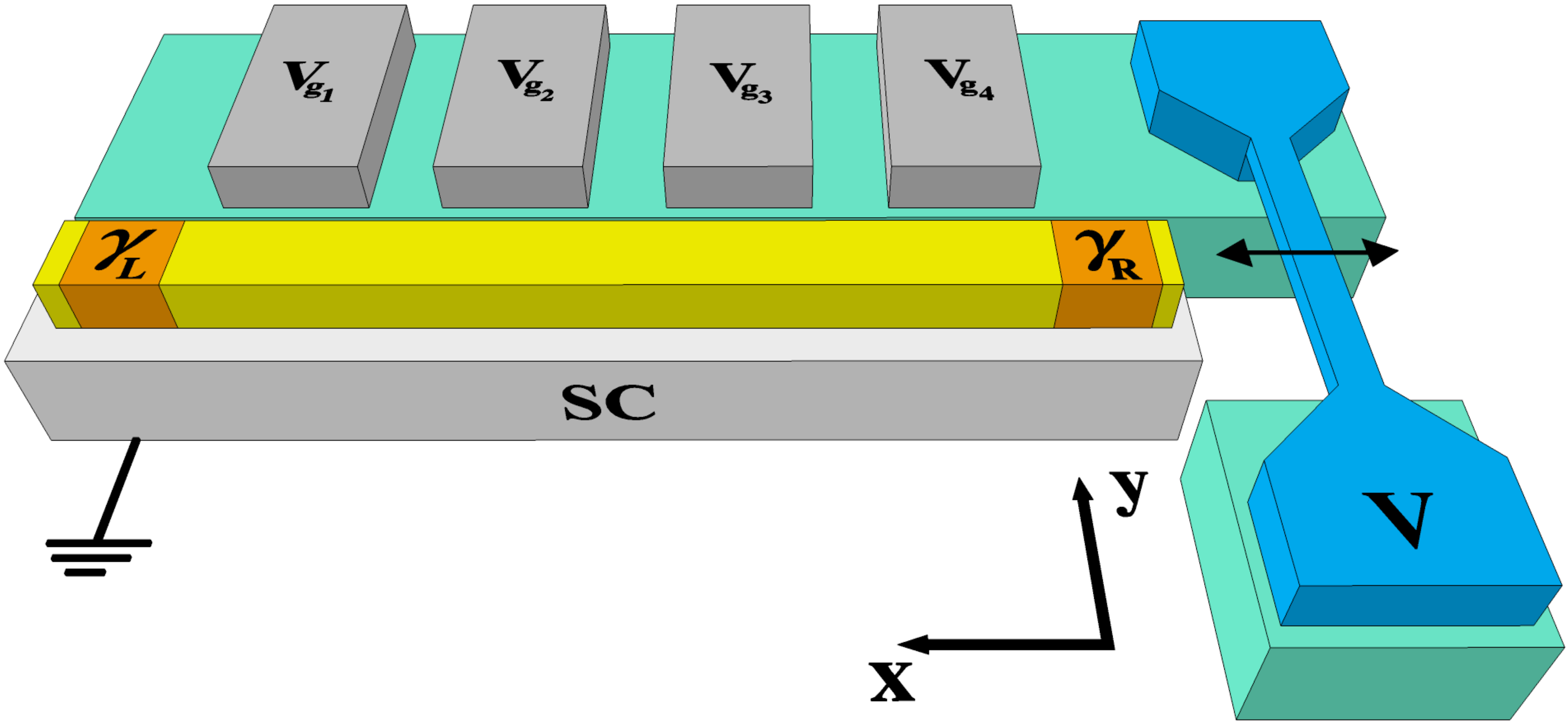}}
	\caption{\label{fig:modelsystem}(Color online)
	Schematic of the setup. A TS is assumed to be realized as a 1D semiconducting wire on top of a grounded $s$-wave superconductor
	(SC). The wire can host MBS at its left ($\gamma_{L}$) and right ($\gamma_{R}$) edges. We assume that one of the edges is tunnel
	coupled to a movable, doubly clamped beam (at bias voltage $V$). The gate electrodes ($V_{g_1}$--$V_{g_4}$) can be used to increase
	or decrease the overlap $\xi$ of the MBS by changing the effective length $L$ of the TS.}
\end{figure}

The detection scheme we propose is schematically depicted in Fig.~\ref{fig:modelsystem}. The TS can, for instance, be realized as
a semiconducting wire with strong spin-orbit coupling placed on top of an $s$-wave superconductor in the presence of a magnetic
field \cite{Sau:2010p254,Lutchyn:2010p077001,Oreg:2010p177002}. Its left and right edges host two MBS which we call $\gamma_{L}$
and $\gamma_{R}$. One of the edges, say the right one, is coupled by tunneling to a movable doubly clamped beam. While this setup
can be technically challenging to realize, one should note that the fabrication of metallic doubly clamped nanoresonators with very high
frequencies and quality factors has been recently reported \cite{FlowersJacobs:2007p182,Li:2008p043112}. The motion of the beam
modulates the tunnel amplitude, such that it depends on the displacement $\hat{x}$ of the beam. The oscillating electrode is held at a
bias voltage $V$ and the TS is grounded.

We approximate the tunnel amplitude as linear in the displacement of the beam, $\at - \bt \hat{x}$. This approximation is justified
if the oscillation amplitude is small compared to the mean distance between the beam and the edge of the TS. The Hamiltonian
of the system is then given by
\begin{align}\label{eqn:ha9}
	H = H_{\rm{res}} + H_{\rm{osc}} + H_{\rm{MBS}} + H_{\rm{tun,0}} + H_{\rm{tun,x}} \, ,
\end{align}
where $H_{\rm{res}} = \sum_{k} \varepsilon(k) \, \psi_{k}^{\dag} \, \psi_{k} $ describes a spinless electron reservoir in the metallic oscillating electrode.
Here, we restrict ourselves to one spin channel only. This is appropriate since the one-dimensional (1D) semiconducting wire on top of an $s$-wave
superconductor in the presence of a magnetic field has the same degrees of freedom as a spinless $p_{x} + i p_{y}$ superconductor
\cite{Fu:2008p273,Oreg:2010p177002}. In reality, both spins will participate and their signal will add up to the one we calculate below.
Furthermore, we assume a linear dispersion $\varepsilon(k)$. Next, $H_{\rm{osc}} =  \hat{p}^{2}/(2m) + m\Omega^{2}\hat{x}^{2}/2$ is the
usual harmonic oscillator Hamiltonian describing the motion of the beam with an effective mass $m$ and resonance frequency $\Omega$.
The term $H_{\rm{MBS}} = i \xi \gamma_{L} \gamma_{R}/2$ characterizes the overlap between the MBS on the left edge and the right edge
of the TS \cite{Semenoff:2007p1479}. Describing these two Majorana fermions as one Dirac fermion, we can rewrite $H_{\rm{MBS}} = \xi  c^{\dag} c$.
The Hamiltonian $H_{\rm{MBS}}$ is formally identical to that of a single resonant level (RL) at energy $\xi$. However, due to the nonlocal nature
of the MBS, the overlap $\xi$ depends exponentially on the effective length $L$ of the TS. This distinguishes the MBS Hamiltonian from a RL.
In order to probe the length dependence of $\xi$, gate electrodes can be installed in proximity to the TS \cite{Hassler:2010p259,Alicea:2011p260},
schematically shown in Fig.~\ref{fig:modelsystem}. Then, by applying gate voltages, the effective length $L$, and thus $\xi$, can be tuned.

The bare tunnel Hamiltonian $H_{\rm{tun,0}}$ has been introduced for a related setup in Ref.~\onlinecite{Bolech:2007p237002} and the
$\hat{x}$-dependent term $H_{\rm{tun,x}}$ is new here. Assuming that the tunneling takes place locally at $y = 0$, both are given by
\begin{align}\label{eqn:ha44}
	H_{\rm{tun,0}} &=  \at [\psi^\dag(y=0) - \psi(y=0)] \gamma_R , \\
	H_{\rm{tun,x}} &=  - \bt \hat{x} [\psi^\dag(y=0) - \psi(y=0)] \gamma_R. \label{eqn:Htunx}
\end{align}
Note that, due to the form of $H_{\rm{tun,0}}$ and $H_{\rm{tun,x}}$, the Majorana fermion couples to lead electrons as well as lead holes.
We cast Eq.~(\ref{eqn:ha44}) into a form containing Dirac fermions:
\begin{align*}
	H_{\rm{tun,0}} &= \at \left\{ \left[ c^{\dag} \psi(0) + \psi^{\dag}(0) c \right] + \left[ \psi^{\dag}(0) c^{\dag} + c \psi(0) \right] \right\} \, ,
\end{align*}
and analogously for $H_{\rm{tun,x}}$. We assume that the lead electrons only couple to $\gamma_{R}$ in $H_{\rm{tun,0}}$ and $H_{\rm{tun,x}}$,
which is justified if the wire is much longer than the superconducting coherence length $\xi_{\rm{SC}}$ of the $s$-wave SC, the
characteristic localization length of the MBS\cite{Fu:2008p273}.

Since MBS are subgap states, we investigate the regime of a large superconducting gap $\Delta_{\rm{SC}}$, i.e., $\rho_{0} t_{0}^{2}, \xi, eV \ll \Delta_{\rm{SC}}$,
where $\rho_{0}$ is the density of states of the metallic lead.

If the electrode does not oscillate, the system is described by the quadratic Hamiltonian $H_{0}=H_{\rm{res}} + H_{\rm osc} + H_{\rm{MBS}} + H_{\rm{tun,0}}$.
In this case, all Green's functions (GF) involving $\psi$ and $c$ operators are known exactly. Since $H_0$ does not conserve fermion numbers, the anomalous
GF do not vanish, e.g., $\ex{c(t) c(0)}_{0} \neq 0$. Then, nonequilibrium transport properties like the current or the current noise can be determined exactly \cite{Bolech:2007p237002}.

\section{Current calculation}
\label{sec:current}

We shall treat the $\hat{x}$-dependence of the tunneling amplitude as a perturbation to the Hamiltonian $H_{0}$. This requires
that $t_{x} x_{\rm{zpf}} \sqrt{2 \ex{n} + 1} \ll t_{0}$,  where $x_{\rm{zpf}}=1/\sqrt{2 m \Omega}$ is the amplitude of the zero point
fluctuations and $\ex{n}$ the mean phonon number of the resonator. We restrict ourselves to second order perturbation theory
in $H_{\rm{tun,x}}$. Our main focus is the calculation of the current. Putting $\hbar = 1$, the current operator is given by
$I = -e \, \partial_{t} N = -ie [H,N]$, where $N = \int dy \psi^\dag(y) \psi(y)$ denotes the number of fermions in the lead.
Using the vector notation $\vec{\Psi} = \left( c, c^{\dag}, \psi(0), \psi^{\dag}(0) \right)$, we find $I = I_0 + I_{\rm{x}}$, where
\begin{align}
	I_0 &= -ie \at \vec{\Psi}^T \mathbf{B}\vec{\Psi} \, , \\
	I_{\rm{x}} &= ie \bt \hat{x} \vec{\Psi}^T \mathbf{B} \vec{\Psi} \, .
\end{align}
Here, $\mathbf{B}$ is a real $4\times4$ matrix
\begin{align}
	\mathbf{B} = \left(\begin{array}{cccc}0 & 0 & 1 & 0 \\0 & 0 & 1 & 0 \\0 & 0 & 0 & 0 \\-1 & -1 & 0 & 0\end{array}\right) \, .
\end{align}
We introduce the fermion GF
\begin{align}
G_{\Psi_{j} \Psi_{k}}(t,t') \equiv G_{jk}(t,t') = -i \ex{T_{C} \, \Psi_{j}(t) \Psi_{k}(t')}_{0}
\end{align}
where $j,k \in \{1,2,3,4\}$ and $T_{C}$ denotes the time ordering operator on the Keldysh contour. Later we will introduce Kelysh indices
$\{-,+\}$ referring to the lower (time-ordered) and upper (anti-time-ordered) branch of the Keldysh contour, respectively. The average is taken with
respect to the ground state of the unperturbed Hamiltonian $H_0$. For $\bt = 0$, the time-independent average current can be
written as
\begin{align}\label{eqn:I0}
	\ex{I_0} = -\frac{e \at}{2} \int \frac{d\omega}{2\pi}  \, \Big\{ &[ G^{K}_{\psi c^{\dag}}(\omega) - G^{K}_{c \psi^{\dag}}(\omega) ] \nn \\
							&+[ G^{K}_{\psi^{\dag} c^{\dag}}(\omega) - G^{K}_{\psi c}(\omega) ] \Big\} \, ,
\end{align}
where
\begin{align}
	G^K_{\Psi_{j} \Psi_{k}}(t,t') = -i \ex{ [ \Psi_j(t), \Psi_k(t') ]}_{0}
\end{align}
is the Keldysh Green's function, for which in general we have $G^{K}(t,t') = G^{-+}(t,t') + G^{+-}(t,t')$.

Next, we consider the corrections to this current for small $\bt$. Using a similar matrix notation allows us to write
\begin{align}
	H_{\rm{tun,x}} = -\bt \hat{x} \vec{\Psi}^T \mathbf{A} \vec{\Psi} \, ,
\end{align}
with
\begin{align}
	\mathbf{A} = \left(\begin{array}{cccc}0 & 0 & 1 & 0 \\0 & 0 & 1 & 0 \\0 & 0 & 0 & 0 \\1 & 1 & 0 & 0\end{array}\right) \, .
\end{align}
Introducing the unperturbed oscillator Green's function
\begin{align}
	D(t,t') = -i \ex{T_C \hat{x}(t) \hat{x}(t')}_{0} \, ,
\end{align}
the first order correction to the fermion Green's function can be expressed as (for $j,k \in \{1,2,3,4\}$)
\begin{align}\label{eqn:gff2}
	&i G^{(1)\alpha\beta}_{jk}(t,t') = \bt \int_{-\infty}^{\infty} d\tau_{1} \sum_{\gamma=\pm} (-\gamma) \nn \\
	& \times \sum_{m,n=1}^{4} \tilde{A}_{mn} \, D^{\alpha \gamma}(t,\tau_{1}) \, G^{\alpha \gamma}_{jm}(t,\tau_{1}) \, G^{\gamma \beta}_{nk}(\tau_{1},t') \, ,
\end{align}
where $\tilde{\mathbf{A}} = \mathbf{A} - \mathbf{A}^{\rm{T}}$ and $\alpha,\beta,\gamma \in \{ -, +\}$ denote the branches of the Keldysh contour.

We express the second-order correction to the GF matrix in terms of the advanced, retarded, and Keldysh components as
\begin{align}\label{eqn:gf9}
	\mathbf{\tilde{G}}_{jk}(t,t') = \mathbf{U} \, \mathbf{G}_{jk}(t,t') \, \mathbf{U}^{\dag} = \left(\begin{array}{cc} 0& G_{jk}^{A} \\  G_{jk}^{R} & G_{jk}^{K} \end{array}\right) (t,t')\, ,
\end{align}
where the transformation is given by
\begin{align}
	\mathbf{U} = \frac{1}{\sqrt{2}} \left(\begin{array}{cc}1 & -1 \\1 & 1\end{array}\right) \, .
\end{align}
This leads to the following compact form of the second-order correction to the GF in the rotated Keldysh space:
\begin{align}\label{eqn:gf10}
	&i \mathbf{\tilde{G}}^{(2)}_{jk}(t,t') \\
	& =- \bt^{2} \int_{-\infty}^{\infty} d\tau_{1} \, d\tau_{2} \, \left[ \mathbf{\tilde{G}}(t,\tau_{1}) \mathbf{\tilde{A}} \mathbf{\tilde{\Sigma}}(\tau_{1},\tau_{2}) \mathbf{\tilde{A}} \mathbf{\tilde{G}}(\tau_{2},t') \right]_{jk} \, ,\nn 
\end{align}
where the effects of the oscillator are contained in the self-energy
\begin{align}\label{eqn:gf14}
	 & \mathbf{\tilde{\Sigma}}_{jk}(t,t') = \mathbf{U} \mathbf{\Sigma}_{jk}(t,t') \mathbf{U}^{\dag} = \\
	 & \frac{1}{2} \left(\begin{array}{cc} D^{A}G^{A}_{jk} + D^{R} G^{R}_{jk} + D^{K} G^{K}_{jk} & D^{R} G^{K}_{jk} + D^{K} G^{R}_{jk} \\ D^{K} G^{A}_{jk} + D^{A} G^{K}_{jk} & 0 \end{array}\right)  \nn \, .
\end{align}
We will assume that the mechanical oscillator has a very high quality factor, such that the linewidth of the uncoupled oscillator is small compared to the effective
linewidth of the fermionic level at $\xi$. With these assumptions, we can use the following advanced, retarded and Keldysh GF in Fourier space:
\begin{align}
	D^{R}(\omega) &= i \pi [\delta(\omega+\Omega) - \delta(\omega-\Omega) ]/[2 m \Omega] \, ,\\
	D^{K}(\omega) &= - i \pi \ex{\bar{x}^{2}} \left[ \delta(\omega-\Omega) + \delta(\omega+\Omega) \right] \, ,\\
	D^{A}(\omega) &= [D^{R}(\omega)]^{*} \, ,
\end{align}
with $\bar{x}^{2} = \hat{x}^{2} + \hat{p}^{2}/(m^{2} \Omega^{2})$. Finally, the average current including the oscillator can be written as
$\ex{I} = \ex{I_{0}} + \ex{I_{\rm{x,1}}} + \ex{I_{\rm{x,2}}}$, where $\ex{I_0}$ is given by Eq.~(\ref{eqn:I0}) and the two remaining terms
are ($j \in \{1,2\}$)
\begin{align}
	\ex{I_{\rm{x,j}}} &= \frac{e}{2} (-\at \delta_{2,j} + \bt \delta_{1,j}) \int \frac{d\omega}{2\pi} \\ 
	\times \Big\{ &[ G^{(j)K}_{\psi c^{\dag}}(\omega) - G^{(j)K}_{c \psi^{\dag}}(\omega) ]
    + [ G^{(j)K}_{\psi^{\dag} c^{\dag}}(\omega) - G^{(j)K}_{\psi c}(\omega) ] \Big\} \nn \, .
\end{align}
The analytic result for $\ex{I}$ is too long to be displayed here. We demonstrate in the following that the average current
contains unique information about the MBS which can be most easily identified in the nonlinear differential conductance
$d\ex{I}/dV$.

\section{Results}
\label{sec:results}

Compared to earlier proposals \cite{Bolech:2007p237002}, we have two additional energy scales involved in the detection
scheme: the resonance frequency $\Omega$ of the oscillator and its effective temperature $T_{\rm{eff}}$. Both are to some
extent experimentally tunable. Assuming that the oscillator is in a thermal state, one has $\ex{\bar{x}^{2}}_{th} = 2 x^{2}_{\rm{zpf}} (2 \langle n \rangle + 1)$
where $\ex{n} = 1/\left[\text{exp}(\Omega/ T_{\rm{eff}}) - 1 \right]$ is the mean phonon number of the oscillator and where we set $k_{\rm{B}}=1$.
In the following, we will discuss the regime where $\ex{n}$ is small, i.e., comparable to 1. This is challenging to realize experimentally.
Note, however, that for the highest resonance frequencies of the doubly clamped beams in Ref.~\onlinecite{Li:2008p043112}
($\sim 500$ MHz), the thermal occupation number could actually become less than 1 at typical dilution refrigerator temperatures.
At higher temperatures or lower frequencies, one would have to implement additional cooling schemes to bring the oscillator
to the quantum regime.
In our model, we make the assumption that the electronic system is at zero temperature, $T_{el} = 0$. The side peaks
in the differential conductance can be resolved as long as $T_{el} \ll \Omega$. If the electronic system and the oscillator are
in equilibrium with the same heat bath, this translates to the requirement $\ex{n} \ll 1$. However, $T_{el} \ll \Omega$ can also be fulfilled for larger
$\ex{n}$ if the oscillator is put into an excited state using experimental techniques as, for instance, done in Ref.~\onlinecite{Oconnell:2010p97}.

\subsection{Reviewing the case without the resonator}

\begin{figure}[ht]
	\center{\includegraphics[width=1\columnwidth]{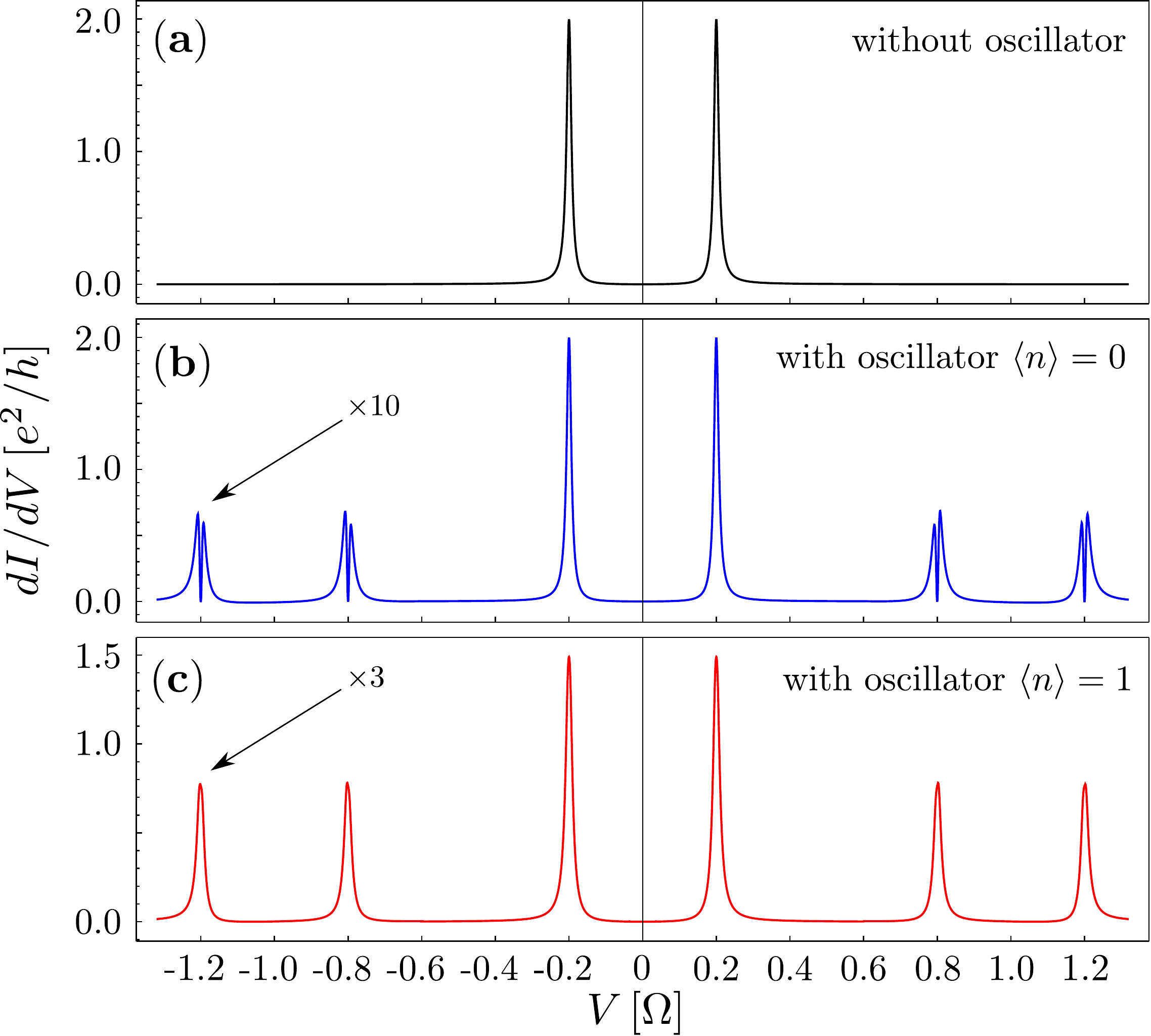}}
	\caption{\label{fig:dIdV1}(Color online) Differential conductance for $\xi/\Omega=0.2$ in the presence of MBS for the case
	$\bt=0$ (without oscillator) as well as for the case $\bt\neq0$ for different values of $\ex{n}$. The satellite peaks have been
	enlarged for better visibility.}
\end{figure}

For clarity, let us start with the case without oscillator ($\bt=0$) and briefly discuss the dependence of the differential conductance
on the length of the TS and thus its dependence on $\xi$. In the case of a finite overlap of the two MBS, the differential conductance
shows two peaks at $eV=\pm|\xi|$ which can be seen in Fig.~\ref{fig:dIdV1} (a) (see also Ref.~\onlinecite{Stanescu:2011p264} for comparison).
Both of the peaks have the height $2 e^{2}/h$ \cite{fnote}.
Importantly, for any finite $\xi$, $d\ex{I}/dV$ at $V=0$ vanishes in our model. This is in stark contrast to the differential conductance
in quantum dots coupled to one superconducting and one normal lead \cite{Domanski:2008p073105}. Hence, it is rather straightforward
to distinguish this situation from the MBS case.
As mentioned above, the Hamiltonian $H_{\rm{MBS}}$ is formally equivalent to one describing a spinless resonant level at energy $\xi$.
Therefore, a comparison to a resonant level case is of pedagogical value.
Two resonant levels with energies $\pm |\xi|$ might lead to a similar signal in the differential conductance as in the MBS case. This could, for
instance, happen if the magnetic field destroyed the superconductivity in an experimental realization of our proposal.
Increasing the effective length of the wire using the gate electrodes, one can tune $\xi$ close to zero, which would yield a single
peak at $V=0$  (not shown in Fig.~\ref{fig:dIdV1}). However, this feature in the differential conductance could also be due to a
single RL with energy $\xi=0$ and consequently an ordinary bound state might mistakenly be identified as a MBS. We conclude
that tunneling into resonant levels could be hard to distinguish from tunneling into MBS.
However, it is fair to say that a measurement of $d\ex{I}/dV$ as a function of the variation of the length of the TS could yield a strong
signature of MBS, even in the case of static leads. Furthermore, $d\ex{I}/dV=2 e^{2}/h$ on resonance in the MBS case,
which might not be the case for tunneling into a RL. Interestingly, we show below that the oscillating electrode exhibits additional
features in the differential conductance, allowing for an unambiguous identification of MBS.
%

\subsection{Differential conductance including the oscillator}

\begin{figure}[ht]
	\center{\includegraphics[width=1\columnwidth]{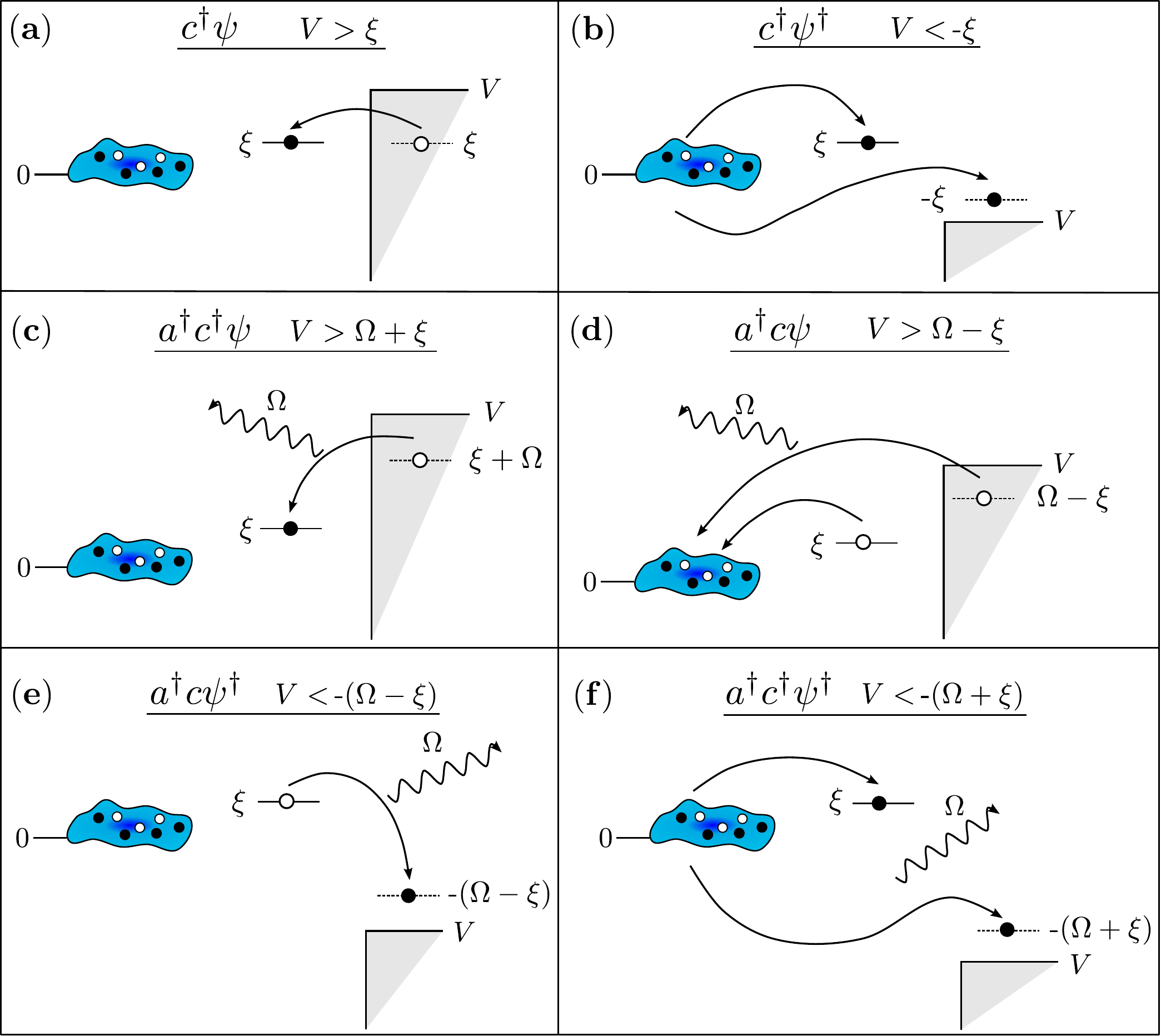}}
	\caption{\label{fig:procs} (Color online) Schematic of energetically allowed tunnel processes for an oscillator in the ground state. Panels a)-f) depict
	the processes stemming from the tunnel terms $c^{\dag} \psi$, $c^{\dag} \psi^{\dag}$, $a^{\dag} c^{\dag} \psi$, $a^{\dag} c \psi$, $a^{\dag} c \psi^{\dag}$, and
	$a^{\dag} c^{\dag} \psi^{\dag}$, respectively, where $\hat{x} = a^\dag + a$. Black dots (circles) represent electrons (holes) and the condensate is depicted as the blue bubble.}
\end{figure}

\begin{figure}[ht]
	\center{\includegraphics[width=1\columnwidth]{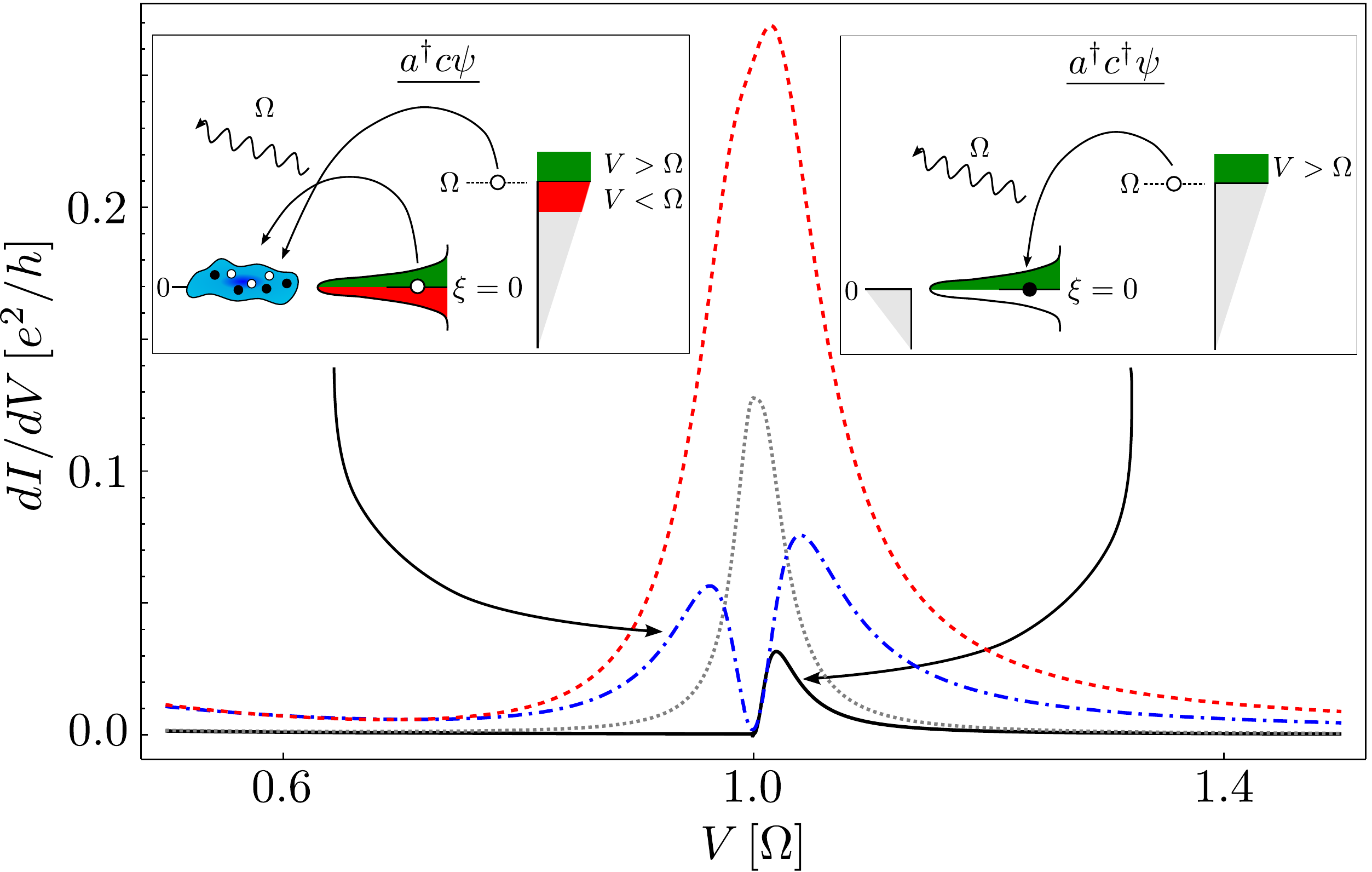}}
	\caption{\label{fig:dIdV2} (Color online) Differential conductance for $\xi\simeq0$ at $eV \approx \Omega$ for a single RL coupled to an oscillator with $\ex{n}=0$
	(solid black line) and $\ex{n}=1$ (dotted gray line). In the former case, the oscillator can enhance transport only for $eV>\Omega$. This is different in the presence
	of MBS. Then, an oscillator with $\ex{n}=0$ gives rise to a positive $d\ex{I}/dV$ for $eV>\Omega$ and $eV<\Omega$ (dashed-dotted blue line). We also show in dashed
	red the $d\ex{I}/dV$ for $\ex{n}=1$ in the MBS situation. The insets depict the decisive tunnel processes for the $\ex{n}=0$ case.}
\end{figure}

The differential conductance in the presence of the oscillator and its dependence on the oscillator's temperature is shown in
Figs.~\ref{fig:dIdV1}(b) and \ref{fig:dIdV1}(c) for two temperatures corresponding to $\ex{n} = 0$ and $1$, respectively. Due to the
presence of the oscillator, satellite peaks emerge at $eV=\pm|\xi\pm\Omega|$. In Fig.~\ref{fig:procs}, we depict the energetically
allowed tunnel processes for $\ex{n}=0$, which explain the emerging satellite peaks in Fig.~\ref{fig:dIdV1}.
(Note that for $\ex{n}=1$ more processes are possible corresponding to the emission of a phonon by the oscillator. These are not shown in Fig.~\ref{fig:procs}.)
Evidently, for the conventional processes in Figs.~\ref{fig:procs}(a), \ref{fig:procs}(c), and \ref{fig:procs}(e), the superconducting condensate does not play any role. Hence,
these processes also matter for tunneling into a RL where the condensate in Fig.~\ref{fig:procs} would be replaced by a
second lead. The processes in Figs.~\ref{fig:procs}(b), \ref{fig:procs}(d), and \ref{fig:procs}(f), on the other hand, rely on the presence of the superconducting condensate.
Importantly, all processes in Figs.~\ref{fig:procs}(a) to \ref{fig:procs}(f) contribute together to the rich structure in the $d\ex{I}/dV$ shown in Fig.~\ref{fig:dIdV1}.
Increasing the oscillator's temperature leads to an increase in the heights of the satellite peaks and transforms a dip at
$eV=\pm|\xi\pm\Omega|$ into a peak. This will be further discussed in Fig.~\ref{fig:dIdV2}.
As mentioned above, one can argue that in a single RL scenario with level energy $\xi=0$, the signal in the differential conductance could not be clearly distinguished from the one
stemming from MBS. Including the oscillator permits us to unambiguously distinguish the two cases. Figure~\ref{fig:dIdV2} shows our key result. In that figure, we compare the
RL case to the MBS case, both in the presence of an oscillating tunnel contact. For clarity, we focus on a region around the resonant frequency $\Omega$ of the oscillator and
positive bias voltage $V$. Importantly, an oscillator in its ground state ($\ex{n}=0$) can only absorb a phonon. In the case of the single RL, the crucial tunnel process near the
resonance at $eV=\Omega$ (for an oscillator with $\ex{n}=0$) is depicted in the right inset of Fig.~\ref{fig:dIdV2}. We see that this process only sets in for voltages $eV>\Omega$
(at zero temperature) since the state in the right lead has to be occupied in order to transfer the energy $\Omega$ to the oscillator. As a consequence, the differential conductance
is only positive for $eV>\Omega$ (shown as a solid black line in Fig.~\ref{fig:dIdV2}). The situation is very different for the MBS case. Here, we have a second crucial tunnel process
depicted in the left inset of Fig.~\ref{fig:dIdV2}. Then, the dashed-dotted blue line in Fig.~\ref{fig:dIdV2} illustrates that, for the oscillator in its ground state and MBS present, the
$d\ex{I}/dV$ is positive for $eV<\Omega$ and $eV>\Omega$. This feature is only due to tunneling through MBS and clearly separates the RL from the MBS scenario.
If we gradually increase $\ex{n}$ from zero to one (where only the two extremes are shown in Fig.~\ref{fig:dIdV2}), the dip for
$\ex{n}=0$ in the MBS case transforms smoothly into a peak due to additional processes where the oscillator emits a phonon.
The dip at $eV=\Omega$ is due to an interference effect between the two participating tunnel processes which can be understood on the basis of Fermi's golden rule.
To illustrate this, we choose $\xi =0$ and $\ex{n} = 0$ and write the system Hamiltonian as $H = \tilde{H}_{0} + \tilde{H}_{\rm{tun}}$ with
\begin{align}
	\tilde{H}_{0} &= H_{\rm{res}} + H_{\rm{osc}} \, ,\\
	\tilde{H}_{\rm{tun}} &= H_{\rm{tun,0}} + H_{\rm{tun,x}}  \, .
\end{align}
In our initial state, the Fermi sea should be filled up to the chemical potential $eV$ (measured from $\xi=0$), furthermore the fermionic subgap state $c$ is empty
and no phonons are present, written as
\begin{align}
	\ket{i} = \ket{ 0_{\psi}, 0_{c}, 0_{a} } \, ,
\end{align}
with energy $E_{i} = E_{F}$, $E_{F}$ being the energy of the filled Fermi sea.

To the lowest order in $\bt$, the leading contribution to $d\ex{I}/dV$ at $eV \approx \Omega$ is generated by a combination of the two tunneling processes $c^\dag \psi$ and $\psi c$,
where one of them is accompanied by the creation of a phonon mode.

The current-carrying transport processes we focus on are mediated by the processes described by $c^{\dag} \psi$ in combination with $\psi c$ (cf. insets in Fig.~\ref{fig:dIdV2}).
For these transport processes the final state lacks two electrons in the Fermi sea (at momenta $|k|,|k'| < k_{F}$), contains one phonon and an unoccupied subgap state $c$, i.e. we have
\begin{align}
	\ket{f} = \psi_{k} \psi_{k'} \ket{ 0_{\psi}, 0_{c}, 1_{a} }
\end{align}
with energy
\begin{align}
	E_{f} = E_{F} - \varepsilon(k) - \varepsilon(k') + \Omega \, ,
\end{align}
where $\varepsilon(k)$ is the single-particle electron energy and the phonon energy is $\Omega$. As usual, the transition amplitude
between the states above, up to second order in $\tilde{H}_{\rm{tun}}$ is given by
\begin{align}
	A_{fi} = \bra{f} \tilde{H}_{\rm{tun}} \frac{1}{E_{i}-\tilde{H}_{0}} \tilde{H}_{\rm{tun}} \ket{i} \, 
\end{align}
and according to Fermi's golden rule, the rate $\Gamma$ is given by summing over all final states respecting energy conservation
\begin{align}
	\Gamma = 2 \pi \sum_{k,k'} |A_{fi}|^{2} \delta(E_{f} - E_{i}) \, .
\end{align}
The rate $\Gamma$ is proportional to the current, and therefore $d\Gamma/d V$ proportional to the differential conductance.
By evaluating the amplitude $A_{fi}$ and the rate $\Gamma$, we find that for the above process, the amplitude $A_{fi}$ vanishes
for tunneling particles with energy close to $eV \approx \Omega$. This leads to the conclusion that for
$eV \approx \Omega$ the differential conductance vanishes due to an interference effect between the two processes (of order $\at \bt$)
generated by the terms $c^{\dag} \psi (\at + \bt a^{\dag})$ and $\psi c (\at + \bt a^{\dag})$ in the tunneling Hamiltonian.
This explains the dip in the differential conductance $d \ex{I}/ dV$ in Fig.~\ref{fig:dIdV2}.

\section{Conclusion}
\label{sec:conclusion}

To conclude, we have presented an idea of coupling Majorana bound states to a sensitive nanoelectromechanical
measurement device. We have shown that a setup where an oscillating, doubly clamped beam is tunnel coupled to a
topological superconductor gives rise to unique transport signatures based on the interplay between the mechanical
excitations and the Majorana bound states. A smoking gun signature of Majorana bound states has been identified
for an oscillator close to the quantum ground state which can be achieved by cooling a $500\,\rm{MHz}$ resonator to
$T\approx 20 \, \rm{mK}$. This energy scale is well below the large parameter of our model, i.e. the superconducting
gap $\Delta_{\rm{SC}}$. When we, for instance, take an $\rm{InAs}$ wire proximity coupled to an $\rm{Nb}$ $s$-wave
superconductor, the superconducting gap of $\rm{Nb}$ $\Delta_{\rm{SC}}^{\rm{Nb}} \sim 15 \, \rm{K}$ can induce a
superconducting gap of $\Delta_{\rm{SC}}^{\rm{InAs}} \sim 1 \, \rm{K}$ in the wire \cite{Lutchyn:2010p077001}. Hence,
our predictions are in principle observable at dilution refrigerator temperatures.

\begin{acknowledgments}
SW and BT acknowledge financial support from the DFG-JST Research Unit {\it Topological Electronics}. TLS acknowledges
support from the Swiss National Science Foundation and KB from the Research Council of Norway, FRINAT Grant 191576/V30.
We would like to thank Jan Budich, Matthew Gilbert, Taylor Hughes, and Ronny Thomale for interesting discussions.
\end{acknowledgments}

\bibliographystyle{apsrev}

\begin{thebibliography}{38}
\bibitem[{Note1()}]{fnote}
\bibinfo{note}{Their height will decrease if we couple the lead
  electrons not only to $\gamma _{R}$ but also to $\gamma _{L}$.}
\end{thebibliography}


\end{document}